\documentclass[pre,twocolumn,showpacs,preprintnumbers,amsmath,amssymb,superscriptaddress]{revtex4}

\usepackage[utf8]{inputenc}

\usepackage{graphicx}
\usepackage{dcolumn}
\usepackage{bm}
\usepackage{colortbl}
\usepackage{epstopdf}

\begin{document}

\title{Complex diffusion-based kinetics of photoluminescence in semiconductor nanoplatelets}

\author{A.~A.~Kurilovich}
\affiliation{Center for Energy Science and Technology, Skolkovo Institute of Science and Technology, 121205, Moscow, Russia}
\author{V.~N.~Mantsevich}
\affiliation{Chair of Semiconductors and Cryoelectronics \& Quantum Technology Center, Physics department, Lomonosov Moscow State University, 119991 Moscow, Russia}
\author{K.~J.~Stevenson}
\affiliation{Center for Energy Science and Technology, Skolkovo Institute of Science and Technology, 121205, Moscow, Russia}
\author{A.~V.~Chechkin}
\affiliation{Institute for Physics \& Astronomy, University of Potsdam, D-14476 Potsdam-Golm, Germany}
\affiliation{Akhiezer Institute for Theoretical Physics 
National Science Center "Kharkov Institute of Physics 
and Technology", 61108, Kharkov, Ukraine}
\author{V.~V.~Palyulin}
\affiliation{Center for Computational and Data-Intensive Science and Engineering, Skolkovo Institute of Science and Technology, 121205, Moscow, Russia}

\date{\today }
\begin{abstract}
We present a diffusion-based simulation and theoretical models for explanation of photoluminescence (PL) emission intensity in semiconductor nanoplatelets. It is shown that the shape of PL intensity curves can be reproduced by the interplay of recombination, diffusion and trapping of excitons. The emission intensity at short times is purely exponential and is defined by recombination. At long times it is governed by the release of excitons from surface traps and is characterized by a power-law tail. We show that the crossover from one limit to another is controlled by diffusion properties. This intermediate region exhibits a rich behaviour depending on the value of diffusivity. Proposed approach reproduces all the features of experimental curves measured for different nanoplatelet systems.
\end{abstract}

 \keywords{} \maketitle

The hunt for materials and systems with better optical properties
has always been one of the focal points of semiconductor research \cite{Ivchenko}. The first classical optoelectronical applications were based mostly on bulk properties of semiconductors. Recent progress in material synthesis \cite{Colvin,Klimov,Nazzal} led to wider research efforts concentrated on low-dimensional structures. In particular, significant attention is drawn by the semiconductor nanocrystals, such as quantum dots (QDs) and nanoplatelets (NPLs)~\cite{Ithurria,Xu}. QDs and NPLs show large exciton binding energy ~\cite{PhysRevB.89.035307,Grim,Achtstein} and possess such remarkable properties as narrow emission lines even at room temperature, tunable emission wavelength, short radiative lifetimes, giant oscillator strength, high quantum yield and vanishing inhomogeneous broadening ~\cite{Ithurria_2,Mahler_1,Tessier,Biadala,Pelton}. These properties make them excellent prospective building blocks for future optoelectronic devices such as bright and flexible light emitters~\cite{She}, colloidal lasers~\cite{Grim,Guzelturk} or for biomedical labeling applications ~\cite{Bruchez}.

Modern techniques of precise colloidal synthesis allow to control the shape~\cite{Manna}, size~\cite{Murray} and crystal
structure of the nanocrystals~\cite{Michalet}. Among these a growing popularity belongs to colloidal QDs and NPLs of CdSe with atomically defined thickness \cite{Ithurria,Smirnov,Smirnov_1} and various shapes such as pure QDs and NPLs (core) as well as composite core-shell QDs and NPLs or core-crown NPLs. The composites are made of two different semiconductor materials, for instance, CdSe and CdS.

The low-dimensionality of all these structures effects in a decisive contribution of a crystal surface to control of optical and electronic properties \cite{Nasilowski}. Specifically, the nanocrystal surfaces contain numerous dangling bonds. The dangling bonds which correspond to the undercoordinated surface atoms rearrange themselves by surface reconstruction or adsorb surfactant ligands \cite{Owen}. The latter are utilised during the colloidal synthesis as they increase the stability of nanocrystals, prevent crystal stacking and screen them from the environment as well as, most importantly, stabilise the surface by saturating dangling bonds \cite{Boles}. As a result ligands influence surface trap states and, consequently, control photoluminescence (PL) properties \cite{Owen_1,Shornikova}. Since the PL signal occurs due to the electron-hole recombination the optical properties of colloidal nanocrystals are directly controlled by different kinds of surface traps (reversible and irreversible).

An example of direct experimental manifestation of the surface trap role in the PL signals is the phenomenon of QDs blinking \cite{Nirmal,Banin}. Most probable mechanisms of this phenomenon were discussed in details in \cite{Kuno,Eli} with time scale-free properties of the blinking process shown to be a signature of complex surface-dependent behaviour of the system.

The NPLs represent a quasi-2D-systems with a large surface to volume ratio and thicknesses down to few atomic layers \cite{Rabouw,Shornikova,Shornikova1}. In \cite{Rabouw} the authors experimentally demonstrate a major effect of temporary charge carrier trapping on the decay dynamics of pure, core-shell and core-crown CdSe NPLs with results matching the previous measurements for NPLs \cite{Olutas,Rabouw_1}. The power-law decay of PL intensity is observed at long times and authors even discerned transitional power laws in some of the cases (for the description of this observation in Ref. \cite{Rabouw} the term multiexponential was used). This type of decay was associated with reversible trapping without non-radiative losses. The power laws changed with temperature rather weakly (the authors drew parallel lines on a log-log plot, i.e. stated that the effect is temperature independent, however, looking carefully one is able to identify some variation with temperature). For blinking from single CdSe QDs the power-law statistics has been found to be independent on the temperature \cite{Kuno,Eli} which points out that blinking and delayed emission may share the same physical origin. Moreover, the model of irreversible charge trapping which leads to the nonradiative recombination was proposed in \cite{Kunneman} to explain the discrepancy between PL and transient absorption measurements on NPLs. Among other structures with nontrivial power-law behavior of PL two-dimensional films of GaTe or GaSe are worth mentioning \cite{Zamudio}.

Currently for description of PL intensity curves most of the theoretical/simulation approaches use kinetic rate equations \cite{Rabouw,Rabouw_1,Zamudio,Shornikova}. The difficulty of this paradigm applied for spatially distributed systems lies in a challenge of interpreting the meaning of the rates from a physical point of view. That makes the experimental estimation and verification of the parameters often impossible. Also simple kinetic rate equations can not produce an explanation for power-law statistics observed in experiments \cite{Rabouw}. In the latter reference an ad-hoc term in their kinetic explanation is used to obtain the power-law. However, the origin or meaning of this term is not elucidated. A few attempts were made to include diffusion of excitons explicitly. For instance, in Refs. \cite{QLi,Wu} CdSe/CdS core/crown nanoplatelets were analysed and it was experimentally demonstrated that exciton localisation efficiency is independent of crown size and increases with photon energy above the band edge, while the localisation time increases with the crown size. To analyse their experimental results authors of Refs. \cite{QLi,Wu} also proposed a theoretical 2D model which considers the competition of in-plane exciton diffusion and selective hole trapping at the core/crown interface. In practice the approach was in a reduction of description to the system of kinetic equations and is not capable of prediction of any non-trivial power-law kinetics. The complexity of exciton dynamics was also found for colloidal halide perovskite nanoplateletes \cite{Weidman}, where the presence of both dark and bright exciton populations was responsible for the complex excitons dynamics.

The experimental observation of exciton diffusion was found in monolayer semiconductors for both freestanding and $SiO_{2}$ supported $WS_{2}$ monolayers \cite{Kulig,Zipfel}. The supporting theoretical work considered non-equilibrium exciton transport in monolayer transition metal dichalcogenides where the interactions between excitons and non-equilibrium phonons were taken into account \cite{Glazov}. In \cite{Olutas} the influence of defects on the spontaneous and stimulated emission performances of solution-processed atomically flat quasi-2D nanoplatelets (NPLs) as a function of their lateral size using colloidal CdSe core NPLs was systematically investigated. It was revealed that the photoluminescence quantum efficiency of these NPLs decreases with increasing lateral size of the colloidal particles while their photoluminescence decay rate accelerates. This strongly suggests that nonradiative channels prevail in the NPL ensembles with platelets of extended lateral size. The effect clearly happens due to the bigger defected NPL subpopulations in NPLs with larger surfaces. Exciton diffusion in 2D metal-halide perovskites was
studied in \cite{NatComm2020}. In this experiment the transition from diffusive to subdiffusive regime is clear which points out to the existence of traps with a power-law distribution of trapping times \cite{ScherMontroll}.

In this paper we draw attention to the important role of diffusive nature of excitons in their free state and a power-law statistics of escape from the traps. Our diffusion-based simulations and theoretical analytically solvable model with rates connected to diffusion/trapping properties explain non-trivial kinetics of carriers in the semiconductor nanoplatelets observed in experiments \cite{Rabouw} and allow clear physical interpretation of the rates.

In the following sections we present our simulation and theoretical models and compare the results with experimental data.

\section{Simulation model and comparison to experiments}

\begin{figure}
\includegraphics[width=8cm]{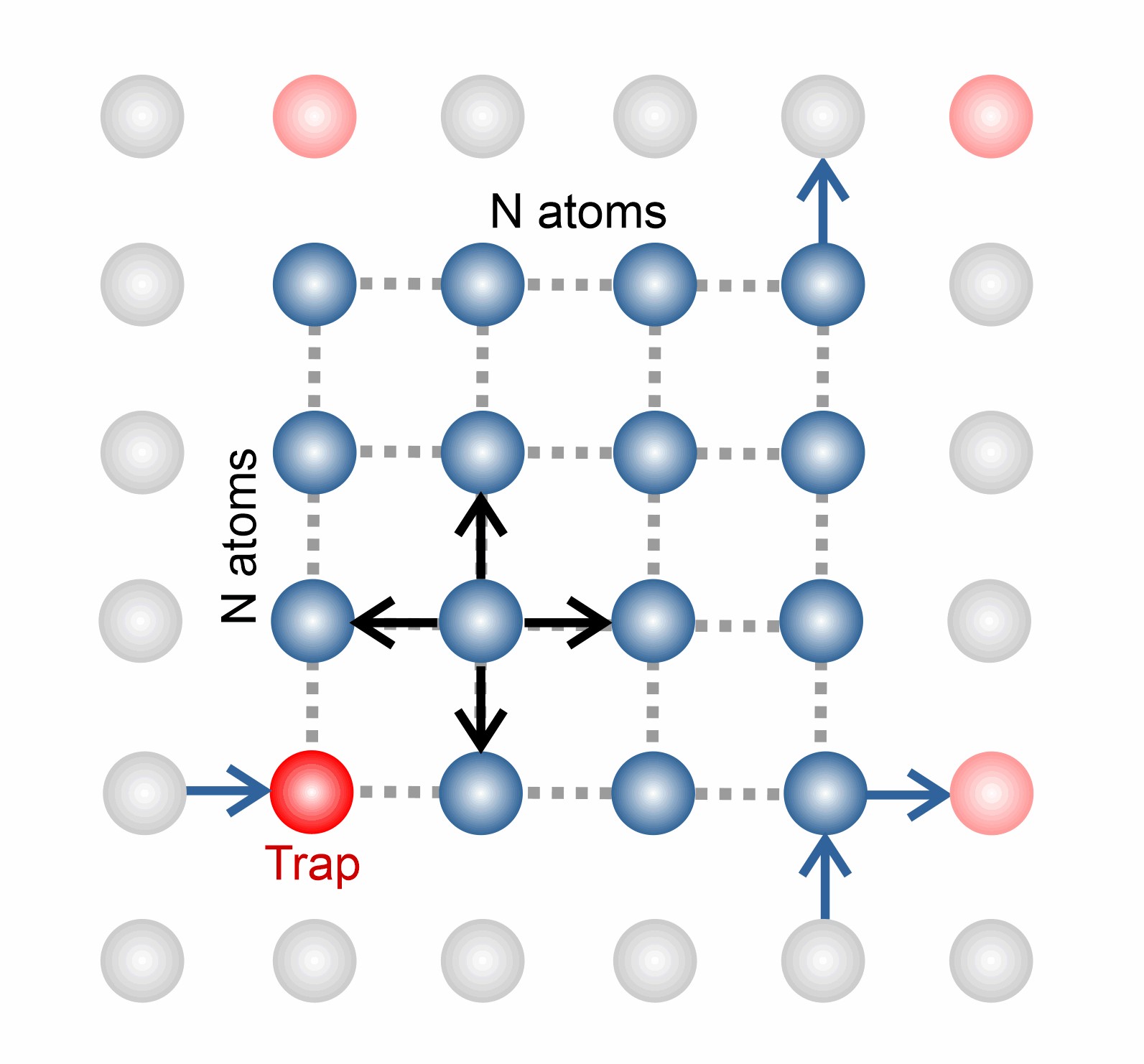}%
\caption{In our simulations we used a model of a random walk on a 2D square lattice modelled as $N\times N$ set of vertices (atoms) with periodic boundary conditions. The jump probabilities are equal in four directions. Some of the vertices are capable of trapping an exciton with a heavy-tail distribution of trapping times $\gamma(t\rightarrow\infty)\sim \frac{1}{t^{1+\mu}},0<\mu<1$. If an exciton jumps on the trap position it gets bound to the cite until it escapes. The recombination is modelled by decay probability $p$ per step.}
\label{fig:Sketch}
\end{figure}

Our simulation model assumes that after the laser beam produces excitons the latter start diffusing on the crystal lattice of a nanoplatelet. During this diffusion process a recombination with a spontaneous photoluminescence or trapping on a surface defect (trap) can happen.  We assume that the recombination of excitons happens only in the free state. Once the particle hits a trap it is considered to be bound (trapped). Eventually excitons leave the traps and again could undergo diffusion, trapping or recombination.

In the model we neglect the possibility of recombination for the trapped excitons. Usually, the trapped PL is well pronounced only for thin nanoplatelets with the width of 1 or 2 monolayers. Its intensity decreases with the growth of number of monolayers. Since we simulated 5 layers which typically corresponds to experiments. Hence, this effect should not be critical for the predictions. Moreover, it is rather weak for the core-shell and core-crown nanoplatelets \cite{Olutas,Saidzhonov}. We also neglect such effects as nonradiative recombination or exciton dissociation. The role of these effects can be controlled experimentally by measuring the quantum yield of the nanoplatelet. The state-of-the-art nanoplatelet synthesis allows to obtain samples with quantum yield varying in the wide range from 10 percent to nearly 100 percent \cite{Erdem}. The role of nonradiative recombination processes and exciton dissociation processes decreases with the growth of the quantum yield. Samples with the quantum yield around 95 percent were synthesised and studied, for example, in Refs. \cite{Zaytsev,Achtstein2,Scott}.

One can estimate the concentration of empty dangling bonds on the nanocrystal surface which are considered to be effective traps following the parameters obtained in \cite{Shornikova}. For CdSe nanoplatelets without shell with the typical largest area size about 100 $nm^{2}$ there exists one paramagnetic center associated with the dangling bond state per 50 Cd surface atoms. In our simulations we choose the values close to that estimate.

\begin{figure}
\includegraphics[width=9cm]{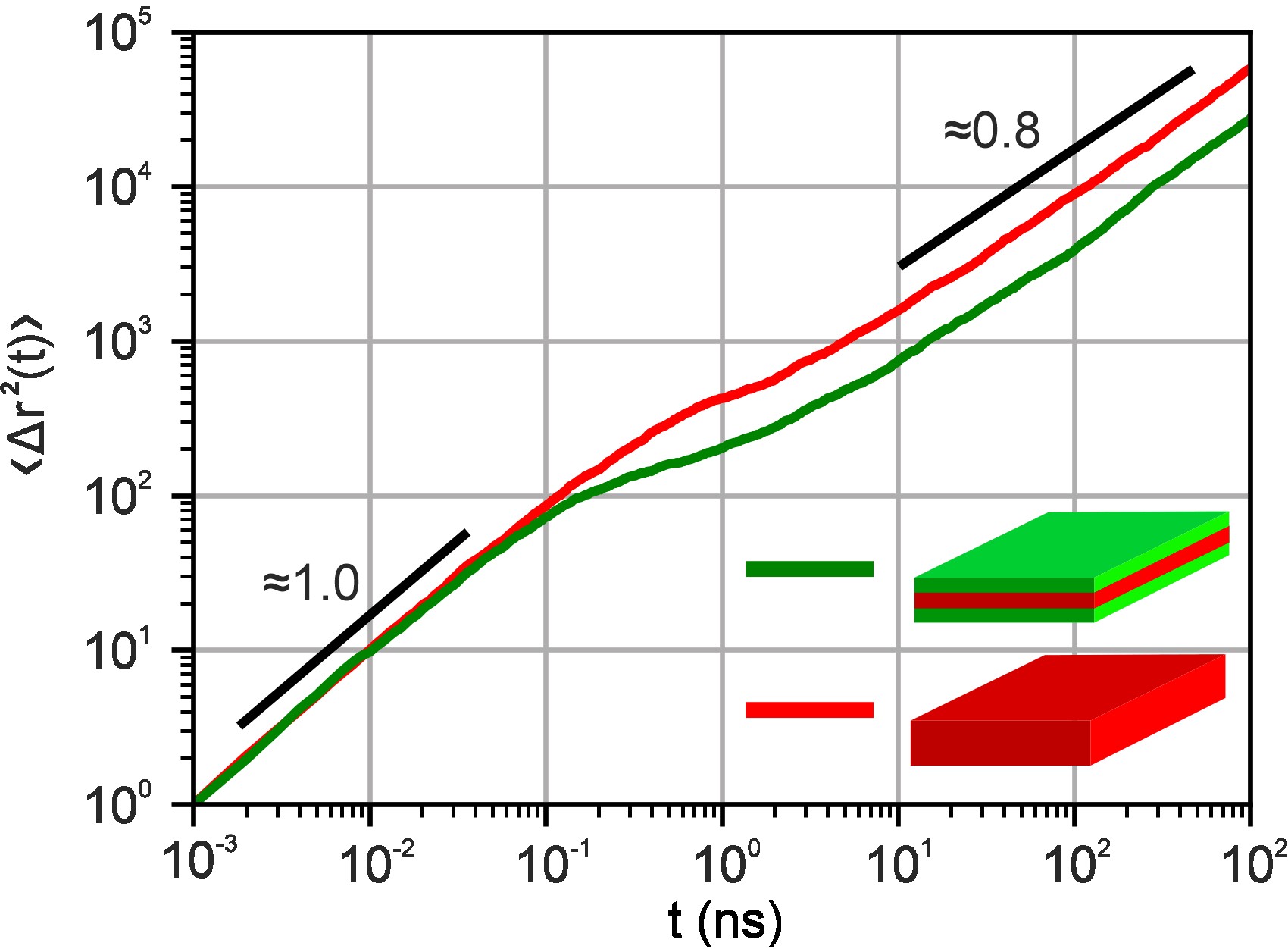}%
\caption{Mean-squared displacement for exciton diffusion as a function of time on double logarithmic scale in the absence of decay. The transition from a normal diffusive regime to subdiffusion is observed. The red curve corresponds to the parameters we use for core structures, i.e.  the simulations were performed on $15\times15$ periodic lattice (i.e.  1 defect per 225 sites), the exponent for trapping times was $\mu=0.8$ and the step duration $\Delta t$=1 ps. The green curve stands for core-shell NPLs and we used $10\times10$ periodic lattice (i.e.  1 defect per 100 sites), the exponent for trapping times was $\mu=0.8$ and the step duration $\Delta t$=1 ps. The curves were obtained by ensemble averaging over 1000 trajectories. The numbers 1.0 and 0.8 show the corresponding slopes.}
\label{MSD}
\end{figure}

\begin{figure}
\includegraphics[width=8cm]{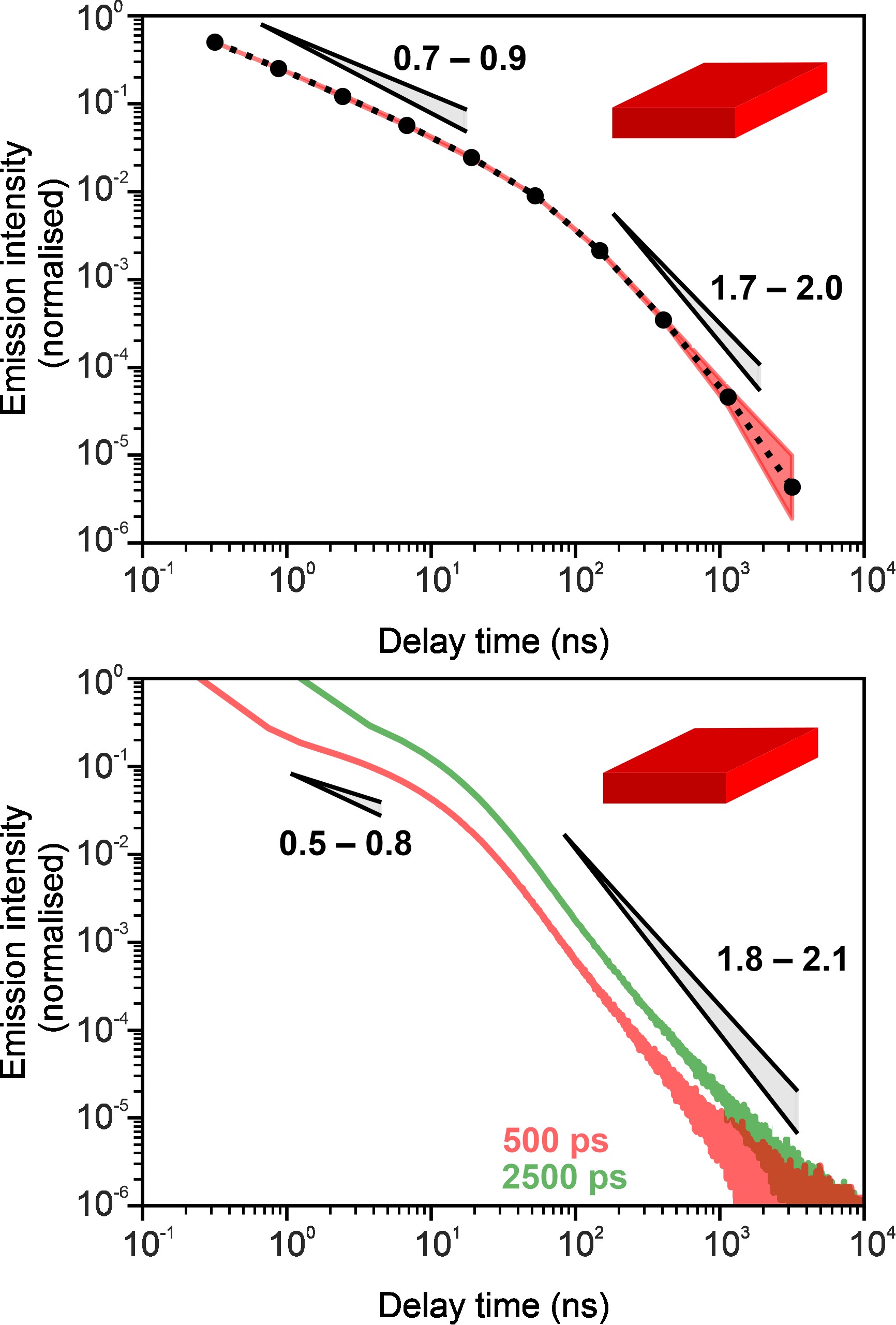}%
\caption{Comparison of digitised experimental data taken from SI of Ref. \cite{Rabouw} (correspond to Fig. 1c in the main text of the reference) (above) with the simulation curves (below) for the case of core CdSe NPLs. The simulations were performed on $15\times 15$ periodic lattice (i.e. 1 defect per 225 sites), the exponent for trapping times was $\mu=0.8$, the step duration $\Delta t=$1 ps and the probability of recombination per step $p=10^{-3}$. $N_{\mathrm{sim}}=10^7$.  Two different binnings are shown in the plot with simulation data. The green curve corresponds to $t_\mathrm{bin}=2500$ ps while the red curve is for $t_\mathrm{bin}=500$ ps. We have recalculated the slopes from Fig. 1c in Ref. \cite{Rabouw}. For both experiment and simulations we show the range of slopes obtained with least mean squares method applied to different subintervals of a seemingly straight part of dependence.} 
\label{SimulationVsExpCore}
\end{figure}

\begin{figure}
\includegraphics[width=8cm]{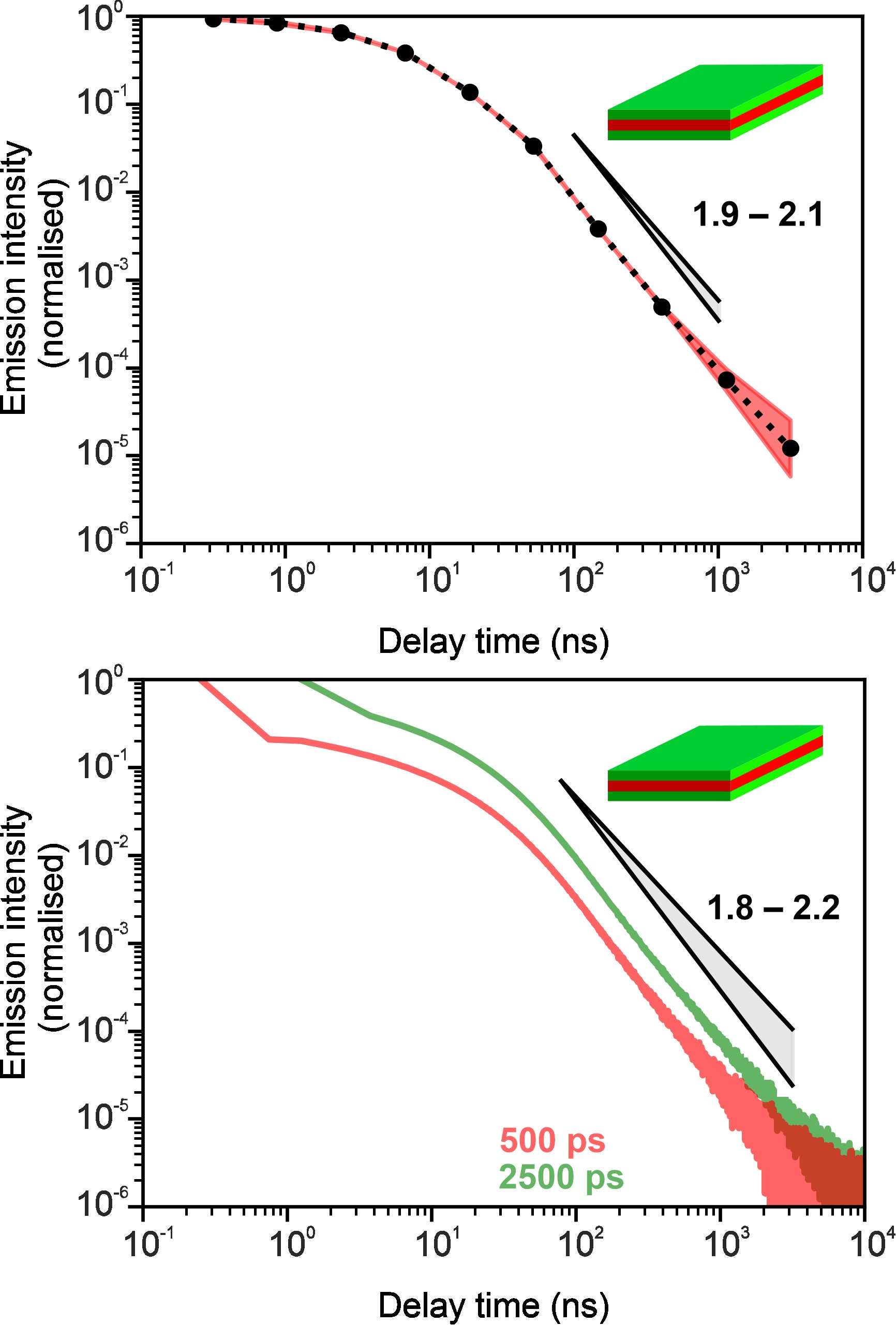}%
\caption{Comparison of digitised experimental data taken from SI of Ref. \cite{Rabouw} (correspond to Fig. 2g in the main text) (above) with the simulation curves (below) for the case of core-shell CdSe-CdS NPLs with CdS forming the outer layers. The simulations were performed on $10\times 10$ periodic lattice (i.e. 1 defect per 100 sites), the exponent for trapping times was $\mu=0.8$, the step duration $\Delta t=$1 ps and the probability of recombination per step $p=10^{-3}$. $N_\mathrm{sim}=10^7$. The green curve in the bottom plot corresponds to $t_\mathrm{bin}=2500$ ps while the red curve is for $t_\mathrm{bin}=500$ ps. We assume that defects are concentrated on the outer surface of CdS. This explains a different selection of trap density in our simulation model for this case (cf. Fig. \ref{SimulationVsExpCore}). We have recalculated the slopes from Fig. 2g in Ref. \cite{Rabouw} and for both experiment and simulations and show the range of slopes obtained with least mean squares method applied to different subintervals of a seemingly straight part of dependence.}
\label{SimulationVsExpCoreShell}
\end{figure}

In order to simulate the exciton diffusion we used a discrete random walk on the 2D square lattice. The finite $N \times N$ lattice with periodic boundary conditions was taken. At every step the simulated particle performs jumps to the nearest neighbours (see Fig. \ref{fig:Sketch}) with equal probability $\frac{1}{4}$. The surface defects (traps) were fixed at the left-down corner of the lattice while the initial exciton position was sampled from the discrete uniform random distribution. The time $\Delta t$ of the jump along the lattice was modelled by a constant value much smaller than the time scale of photoluminescence and trapping processes. The exciton in a free state can recombine with a probability $p$ at each jump, i.e. we neglect any interactions between the excitons which is perfectly justified for low-to-moderate intensities of an initial laser pulse \cite{Irkhin2011}. The trapping occurs with the unit probability when the exciton reaches the trap position. Importantly, in our model the probability distribution of trapping times is non-exponential, having power-law tail such that the mean trapping time is infinite. Such trapping, or waiting time, distributions appear in the theory of transport in disordered media called Continuous Time Random Walk model \cite{MontrollWeiss,ScherLax,ScherMontroll}, see also \cite{Hughes,bouchaud,report2000} and references therein. This distribution of trapping times occurs, for instance, in systems with an exponential distribution of depths of potential wells \cite{tiedje80}. We sample the trapping times from the one-sided $\alpha$-stable distribution with the L\'evy index $\mu, 0 < \mu < 1$ following \cite{chambers}. Such distribution possesses long time asymptotics $\gamma(t) \sim t^{-1-\mu}$. The Laplace transform 
of the distribution reads $\gamma(s) = e^{-\tau_*^\mu s^\mu}$.  In simulations we use $\tau_* = 1$ ns, for more information about $\alpha$-stable distributions see Appendix A. After a particle escapes from the trap it can be trapped again, i.e. the multiple trapping of excitons is taken into account by the model.

As a result the motion of an exciton in our model is characterised by a normal diffusive (Brownian) motion at short times and a subdiffusion with a characteristic power-law exponent $\mu$ at long times. Fig. \ref{MSD} shows the dependence of an exciton mean-squared displacement as a function of time in the absence of decay. Indeed, at first, the particles diffuse freely but then the trapping and a subsequent release become important and the effective motion becomes subdiffusive.

The photoluminescence intensity is defined by the number of exciton recombinations per second, i.e. by the distribution of recombination times. In simulations we obtained this distribution from the set of $N_\mathrm{sim}$ independent runs, i.e. we again use a low exciton density assumption. The normalised emission intensity was calculated by aggregation of the single recombination (photoluminescence) events in the bins of the size $t_\mathrm{bin}$ (corresponding to the time resolution in the experiment \cite{Rabouw}) and further normalisation by the number of events in the first bin in the spirit of experimental work \cite{Rabouw}. The middle of the corresponding bin was used as a time coordinate for the normalised emission intensity. Thus, our $t_\mathrm{bin}$ corresponds to the exposition time in experiment \cite{Rabouw}. Increase of $t_\mathrm{bin}$ leads to the noise reduction on the one hand and to smoothing of the curve and possible loss of fine effects on the other hand. Another important fact is that the highest intensity of PL occurs at short times. Hence by enlarging the bin one gets disproportionately high counts to the first bin. In the latter case we effectively decrease the rest of the values.

In Figures \ref{SimulationVsExpCore},\ref{SimulationVsExpCoreShell} we show the comparison of our simulation model with fitted parameters (bottom plots) with experimental data for the photoluminescence intensity in Ref. \cite{Rabouw} (upper plots). Fig. \ref{SimulationVsExpCore} shows the PL intensity for core CdSe NPLs and Fig. \ref{SimulationVsExpCoreShell} shows the PL intensity for core-shell CdSe-CdS nanoplatelets with CdS forming the outer layers. In order to plot and analyse the experimental data we have used the plots from the Supporting Information for Ref. \cite{Rabouw} and digitised the data with the help of GetData Graph Digitizer program \cite{GetData}. Black dots in the upper plots of Figs. \ref{SimulationVsExpCore}, \ref{SimulationVsExpCoreShell} are the averages of the digitised data with red region around representing the statistical error of measurements. One can see that the power laws including the intermediate ones nicely match between the simulations and experiments. For both the simulation and the experimental Ref. \cite{Rabouw} results we show the range in exponents in order to show how sensitive it is to the size of the interval which visually looks "straight". By showing the results for two different bin sizes in simulations we illustrate that while the long time power-law asymptotics stays the same, the intermediate one is affected by renormalisation. This pushes us to the conclusion that the intermediate power laws are transitive phenomena rather than true power laws. Moreover the adjustment of the bin in simulations could produce a very good quantitative match with experimental data (see Appendix B). However, as we discovered the experimental data consist of two differently normalised parts (see the Supplemental Information for \cite{Rabouw}) and also reveal a range of values for power law asymptotics depending on the time interval chosen for averaging. Hence we focus here on the explanation of the power laws rather than on the attempt to match the data quantitatively.

Note that we were not able to get a good match of the overall curve shapes with experiments for the core-crown case from Ref. \cite{Rabouw} (still the long time power-law tail can be easily reproduced). We believe that the reason is that this case is markedly different. In comparison to pure core or core-shell cases it stands out as a system with a surface divided into two chemically distinctive areas with different properties. This obviously strongly affects the intermediate time scales. Hence, this case would require a more complex modelling approach.

The NPLs studied in experiments normally have a thickness of a few atomic layers. Hence the diffusion into inner layers is also possible. If one takes into account these bulk layers the results stay qualitatively the same which is shown in Appendix C.

\section{Theoretical Model. Non-Markovian kinetic rate equations}

Our theoretical model corresponds to the simulation setup with one simplification. We additionally assume that at every step the trapping happens at a constant rate $\beta$ and is proportional to the number of particles in free state $\beta n_f(t)$, $n_f(t)$ being the number of excitons in the free state as a function of time $t$. This coefficient $\beta$ effectively describes diffusion of excitons and can be connected to properties of the simulation model (see below). The decay (recombination) is assumed to be proportional to the concentration $n_f(t)$ as $\alpha n_f(t)$, where $\alpha$ is a rate of recombination process. Eventually an exciton escapes the trapped state and switches back to the diffusion-decay mode. For the population of excitons in the diffusive state the trapping process presents a temporary loss with a delayed return. This return can be described by the term with a kernel $\gamma(t-\tau)$, $\int_0^t\gamma(t-\tau)\beta n_f(\tau)d\tau$, where $t$ is a release time moment as compared to $\tau$, the time of the trapping event. The kernel has an explicit meaning of trapping probability density as a function of time, i.e. it abides a normalisation property, $\int_0^\infty\gamma(t)dt=1$. Hence, our theoretical model describes the recombination-caused photoluminiscence process as a two-state system model with only one of the states allowing the recombination.

Summarising the above model in terms of kinetic rate equations for concentrations of free excitons $n_f(t)$ and the trapped excitons $n_t(t)$ one gets
\begin{eqnarray}\label{Equations}
&&\frac{dn_f(t)}{dt}=-\alpha n_f(t)-\beta n_f(t)+\int_0^t\gamma(t-\tau)\beta n_f(\tau)d\tau,\\
&&\frac{dn_t(t)}{dt}=\beta n_f(t)-\int_0^t\gamma(t-\tau)\beta n_f(\tau)d\tau.
\label{Equations1}
\end{eqnarray}
The initial conditions are $n_f(0)=N_0, n_t(0)=0$, i.e. the laser excitation produces $N_0$ particles in the free state, but no trapped ones. Applying the Laplace transform $\mathcal{L}\{f(t)\}=\int_0^\infty f(t)e^{-st}dt=\tilde f(s)$ one turns the equations into algebraic ones. The solution for $\tilde n_f(s)$ reads
\begin{eqnarray}\label{exactsolutionfree}
\tilde n_f(s)=\frac{N_0}{s+\alpha+\beta(1-\tilde\gamma(s))}.
\end{eqnarray}
\begin{figure}
\includegraphics[width=8cm]{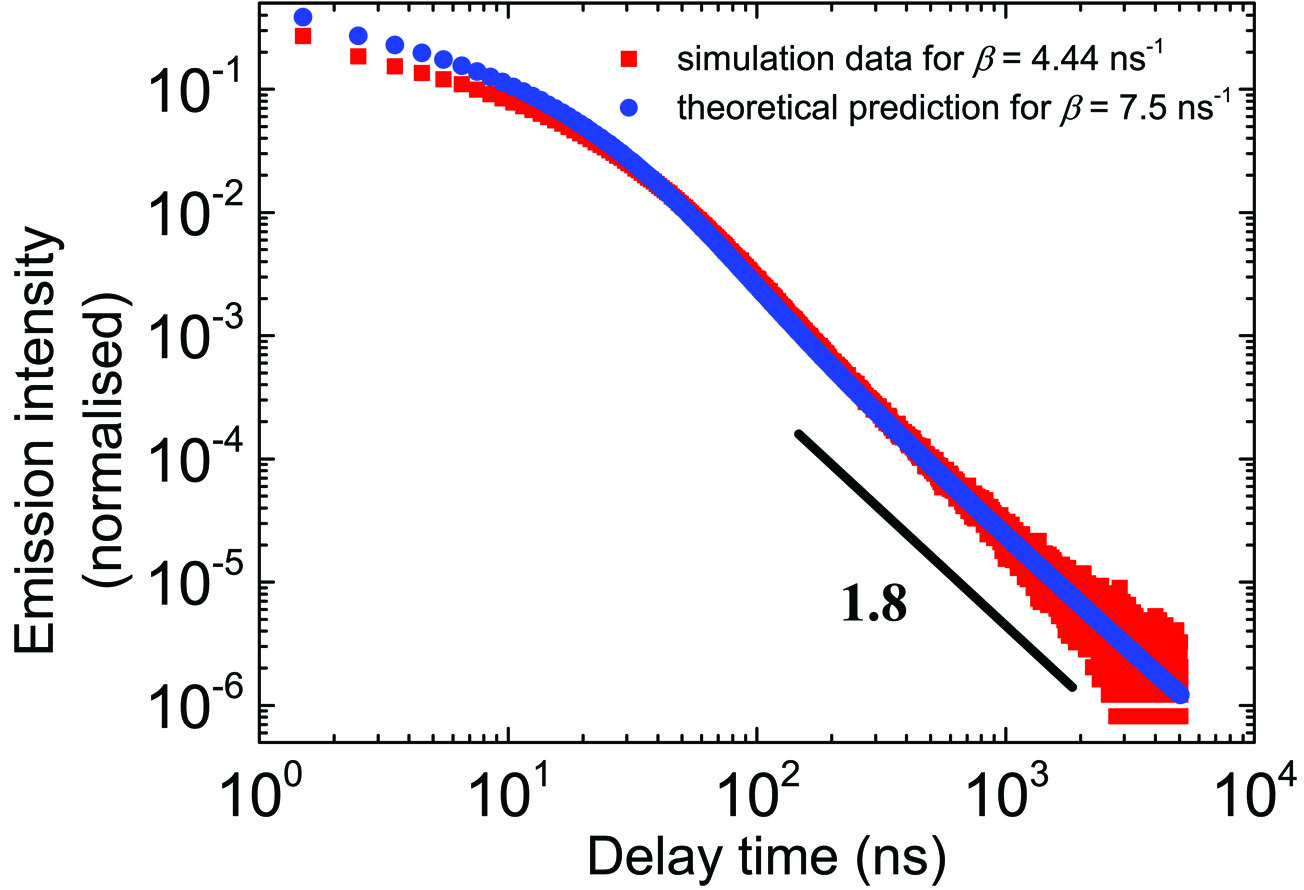}%
\caption{The outcome of the theoretical model vs the simulation results in the double log scale for all relevant time scales. The parameters of theory are $\beta=7.5 \textrm{ns}^{-1},\alpha=1\textrm{ns}^{-1},\mu=0.8$. In simulations $\beta=4.44 \textrm{ns}^{-1}$. Red dots correspond to the simulation data. Blue points are the numerical inverse Laplace transform of Eq. (\ref{exactsolutionfree}). $t_{\mathrm{bin}}=500 \textrm{ps}$}
\label{fig:TheoryVsSimulationsLong}
\end{figure}
Since $\gamma(t)$ is normalised, $1-\tilde\gamma(s)$ never becomes negative and any possible divergencies are avoided. Also it can be proven that the solution $n_f(t)$ is non-negative at all times (Appendix D). Clearly, in a general case the system of Eq. (\ref{Equations}) and (\ref{Equations1}) describes a non-Markovian process due to the memory kernel. The Markovian case can be obtained for the particular case of exponential distribution of trapping times (for more details and the exact solution of that case, see Appendices E,F). We assume that the power law tails observed in experiments appear due to the power law asymptotics of trapping times, i.e. one can use an $\alpha$-stable law for the kernel, $\tilde\gamma(s)=e^{-\tau_*^\mu s^\mu},0<\mu<1$. In the limit of long times $t\rightarrow\infty$ (which corresponds to $s\rightarrow0$) the function $\tilde\gamma(s)\approx1-\tau_*^\mu s^\mu$. Thus, in this limit the concentration of freely diffusing excitons is
\begin{equation}
\tilde n_f(s)\approx \frac{N_0}{\alpha}\left(1-\frac{\tau_*^\mu\beta}{\alpha}s^\mu\right).
\label{limitnf}
\end{equation}
This expression produces a long-time power-law asymptotics $n_f(t\gg\tau_*)\simeq\frac{C \tau^\mu_*}{t^{1+\mu}}$ with $C=\frac{\mu\beta N_0}{\alpha^2\Gamma(1-\mu)}$ (see the derivation in Appendix G). Then the limit expression for the emission intensity is
\begin{equation}
I(t\gg\tau_*)\simeq \frac{\mu\beta N_0}{\alpha\Gamma(1-\mu)}\frac{\tau_*^\mu}{t^{1+\mu}}
\label{intensity}
\end{equation}

In order to compare the theoretical model with the simulations we need to find the correspondence between the simulation parameters and the constants used in the theory. As a test example we use the simulation results from Fig. \ref{SimulationVsExpCore} with the following parameters: the simulation step $\Delta t$ is 1ps, the bin size 500 ps, the time limit of the simulation 10000 ns, 2D square lattice with 1 defect per 15x15=225 points, $\mu=0.8$, probability of recombination per step $10^{-3}$. Hence, the parameters to be put in Eq. (\ref{exactsolutionfree}) can be estimated as follows. For $\tilde\gamma(s)=e^{-\tau_*^\mu s^\mu}$ the coefficients are $\mu=0.8$, $\tau_*=1$ ns. The rate of recombination $\alpha$ can be determined as a probability of recombination per step divided by step's duration, i.e. $\alpha=1 \mathrm{ns}^{-1}$. The rate of exciton capture by defects (traps) $\beta$ is roughly the probability of find a defect per step. This probability equals 1/225 if one neglects the effect of secondary trapping in which case the probability of being caught by the defect is higher after the escape from a defect. For the case of the first trapping event the value of the probability is based on the fact that original excitations occur homogeneously in the sample. For the first trapping case $\beta_{\mathrm{firsttrap}}$ can be then estimated as $\beta_{\mathrm{firsttrap}}=\frac{1}{225}/\Delta t=4.44$ ns$^{-1}$ (the trapping rate is almost constant at short times). In order to account for the undercounting of trapping events at long times in our theoretical model we slightly increase the trapping rate and set $\beta=7.5$ ns$^{-1}$. At long times the secondary trapping is quite likely since right after the escape from the trap a particle is more likely to come back then at earlier times.

Fig. \ref{fig:TheoryVsSimulationsLong} shows the excellent correspondence between the theoretical and simulation results obtained for these parameters. Since the simulation results are normalised on the first bin the theoretical curve was scaled such that the first points roughly coincide.

\begin{figure}
\includegraphics[width=8cm]{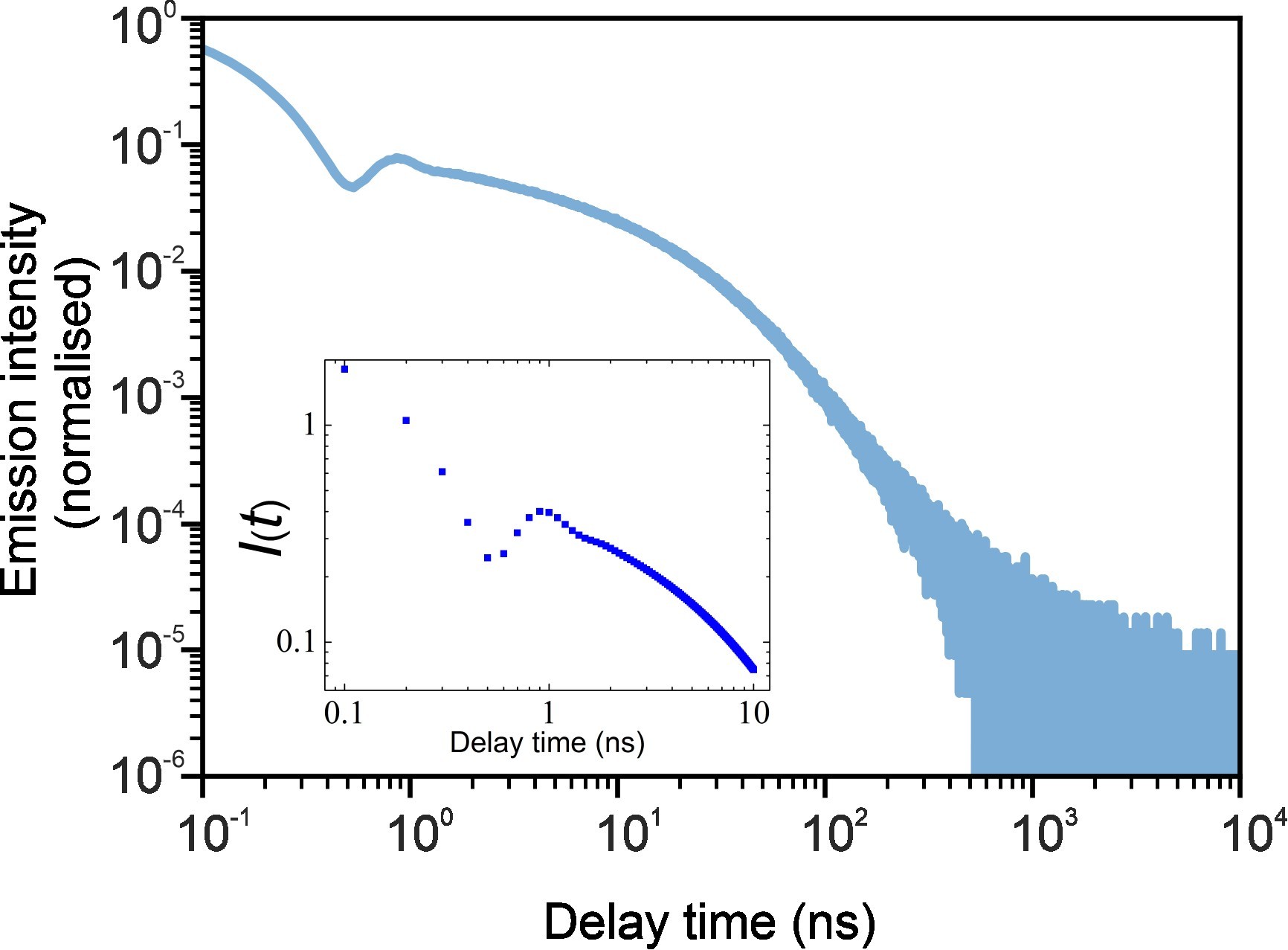}%
\caption{The simulation results (the main plot) with fine binning showing the same intermediate fluctuating behaviour as in theoretical model (shown in the inset). Here the bin size was chosen to be $t_{bin} =$ 25 ps, $\mu=0.8$, $p=10^{-3}, N_{\mathrm{sim}}=10^7$, one trap was put per 100 lattice sites.}
\label{fig:SimulShortTimes}
\end{figure}

\begin{figure}
\includegraphics[width=8cm]{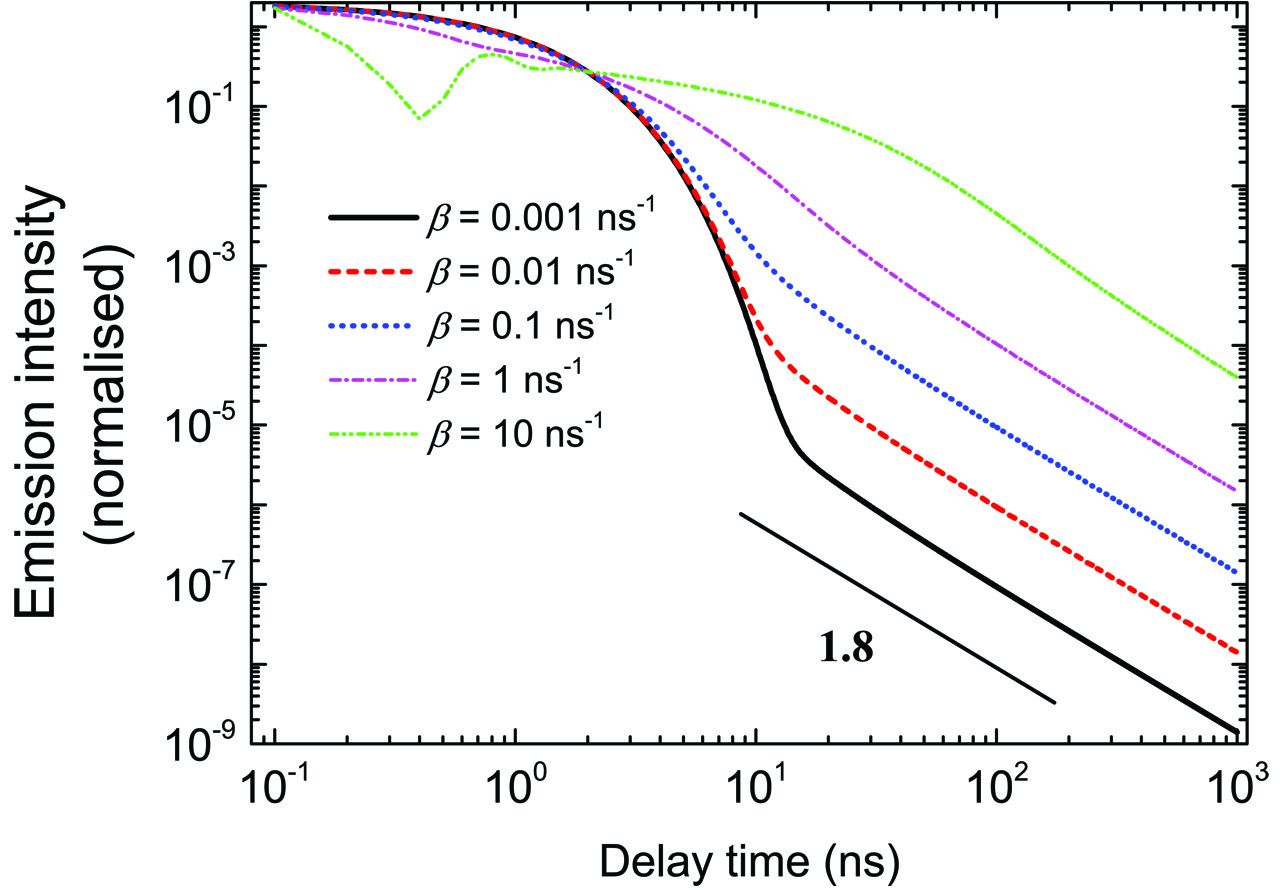}%
\caption{Dependence of logarithm of emission intensity as a function of logarithm of time for different $\beta$. The parameter $\beta$ is defined by the free diffusion coefficient of excitons. The variation of $\beta$ leads to the change in a characteristic shape of the emission intensity curves.}
\label{beta_dependence}
\end{figure}

The study of parameter space both in simulation and theoretical models shows that at the intermediate times a rich behaviour can be observed. In Fig. \ref{fig:SimulShortTimes} for the chosen parameters the curve weekly oscillates which is a result of an interplay of trapping, release and recombination terms both in simulations (the main plot) and in theory (the inset). 

The shapes of the emission intensity curves are defined by the interplay between free diffusion, trapping and recombination of excitons. One can see from Fig. \ref{beta_dependence} that the overall shape is controlled by diffusion properties. For any set of parameters at short times emission is defined by recombination only and the intensity decays exponentially. The latter observation was made experimentally, e.g. see Ref. \cite{Shornikova}. Exponential decay at short times also follows from Eq.(3). At large $s$ which correspond to the short time domain $\tilde n_f(s) \sim N_0/(s+\alpha+\beta)$, that is $n_f(t) \sim N_0 \exp(-(\alpha+\beta)t)$. At short times most of the excitons are free. In the course of time they get trapped and escape the traps later. Since the trapping times are described by a power-law with exponent $\mu+1$ at long times the emission is defined by the release of excitons and characterised by the same power-law. The parameter $\beta$ defines both the duration of the short time limit and the properties of the transition region. At small $\beta$ (slow diffusion) the transition from exponential to power-law decay happens without a noticeable transition region (see the black solid curve in Fig. \ref{beta_dependence}). Then with a growth of $\beta$ the intermediate region appears which is well illustrated by the pink dash-dotted curve in Fig. \ref{beta_dependence}. This is a regime when intermediate power laws were suggested in Ref. \cite{Rabouw}. Finally, for fast diffusion (green dash dot-dotted curve in Fig. \ref{beta_dependence}) one can observe a wide transition region with oscillations of emission intensity. These oscillations could be associated with multiple trapping and release events.

\section{Summary}

We proposed a diffusion-based simulation model and non-Markovian kinetic rate equations with a delay function, which describe the interplay between free diffusion, trapping and recombination of excitons in 2D semiconductor nanostructures. The parameters of both models have clear physical interpretation. The models were applied for the interpretation of experimental data on photoluminescence kinetics in CdSe/CdS core and core-shell nanoplatelets. The characteristic power-law tails are the result of exciton escape properties from surface defect traps. We show that the intermediate power laws found in Ref. \cite{Rabouw} are transitional effects rather than true power-law dependencies. Moreover, we see that diffusion properties control the emission intensity in the intermediate region between the simple exponential decay at short times and a power-law at long times. The duration and the features of this intermediate region vary substantially with the value of diffusivity. In particular, in the case of a fast diffusion complex oscillating dependencies are demonstrated. The proposed model can be applied for the analysis of exciton kinetics in semiconductor nanoplatelets (core, core-shell, core-crown) and for estimation of microscopic exciton parameters. In the Appendix C we show a possible pathway for the model generalisation for the 3D case.  Further generalisations of the proposed model and additional assumptions may be necessary to use the model for other low-dimensional semiconductor structures.

\section*{Conflicts of interest}
There are no conflicts to declare.

\section*{Acknowledgements}
The authors would like to acknowledge the support of Russian Science Foundation project No. 18-72-10002 as well as fruitful discussions with Igor M. Sokolov.

\appendix

\section*{Appendix A: $\alpha$-stable L\'evy laws}\label{alphastable}

In our model we choose a one-sided (completely asymmetric) $\alpha$-stable L\'evy distribution as a model distribution for the statistics of trapping times. The reasons for that are both mathematical and practical. 

Mathematically, according to the generalised central limit theorem the distribution of the properly normalised sum of independent identically distributed variables converges to $\alpha$-stable L\'evy law if the variables are drawn from a distribution with a divergent second moment \cite{gnedenkokolmogorov}. The $\alpha$-stable distributions can only be written analytically in terms of often hard-to-grasp Fox H-functions with three exceptions (Gaussian, Cauchy and L\'evy-Smirnov distributions). However, their Fourier/Laplace transforms adapt a simpler stretched Gaussian/exponential form \cite{samorodnitskii}. In the case of trapping time distributions the quantities adopt the positive values only, i.e. one needs to consider a particular case of one-sided alpha-stable distributions. For the latter the Laplace transform reads
\begin{equation}
G(s)=\exp(-\tau^\mu_* s^\mu),
\end{equation}
where $0<\mu\le1,s\in[0,\infty)$. 
At long times this distribution exhibits a power-law asymptotics which decays (up to a numerical prefactor) as $\frac{\tau_*^\mu}{t^{1+\mu}}$.

The practical reason for our choice is a convenience of working with exponential dependencies both in terms of expansions and numerical calculations. In simulations the distribution was sampled according to Ref. \cite{chambers}.

\section*{Appendix B: Match between theory and experiment for the binning value $t_\mathrm{bin}=2.5$ ns}

\begin{figure}
\includegraphics[width=8cm]{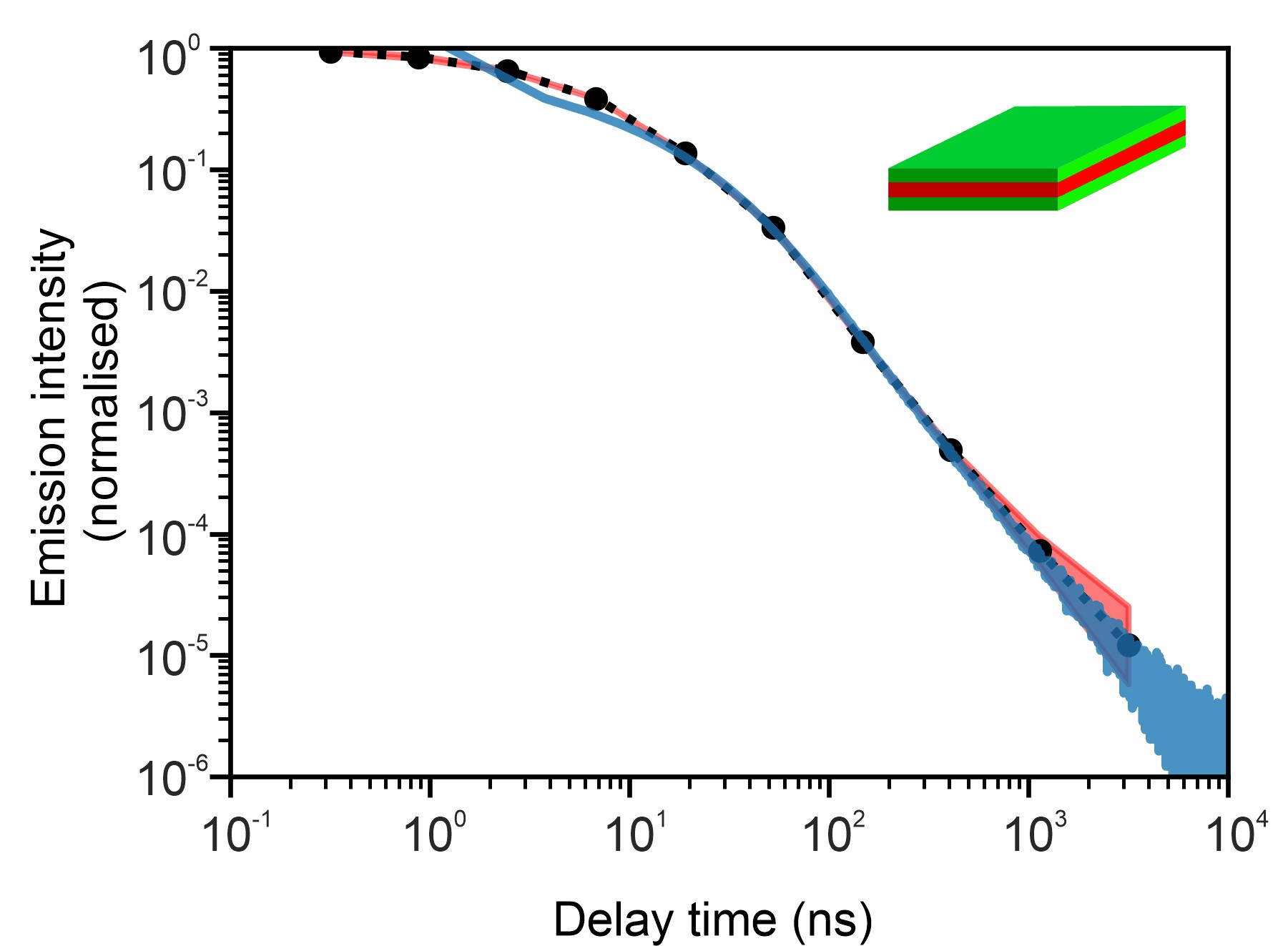}%
\caption{Quantitative match for photoluminescence intensities between the simulations and experiment from Ref. \cite{Rabouw} for the binning $t_\mathrm{bin}=2.5$ ns for the case of core-shell CdSe-CdS NPLs with CdS forming the outer layers.}
\label{match}
\end{figure}

If one plots the simulation and experimental results from Fig. 4 for the binning 2.5 ns = 2500 ps one finds even a very good quantitative match (Fig. \ref{match}). However, one has to keep in mind that the values depend on the normalisation which was not always known from experimental data in Ref. \cite{Rabouw}.

\section*{Appendix C: Modelling the process in a few atomistic layers}\label{Appendix2Dvs3D}

\begin{figure}
\includegraphics[width=8cm]{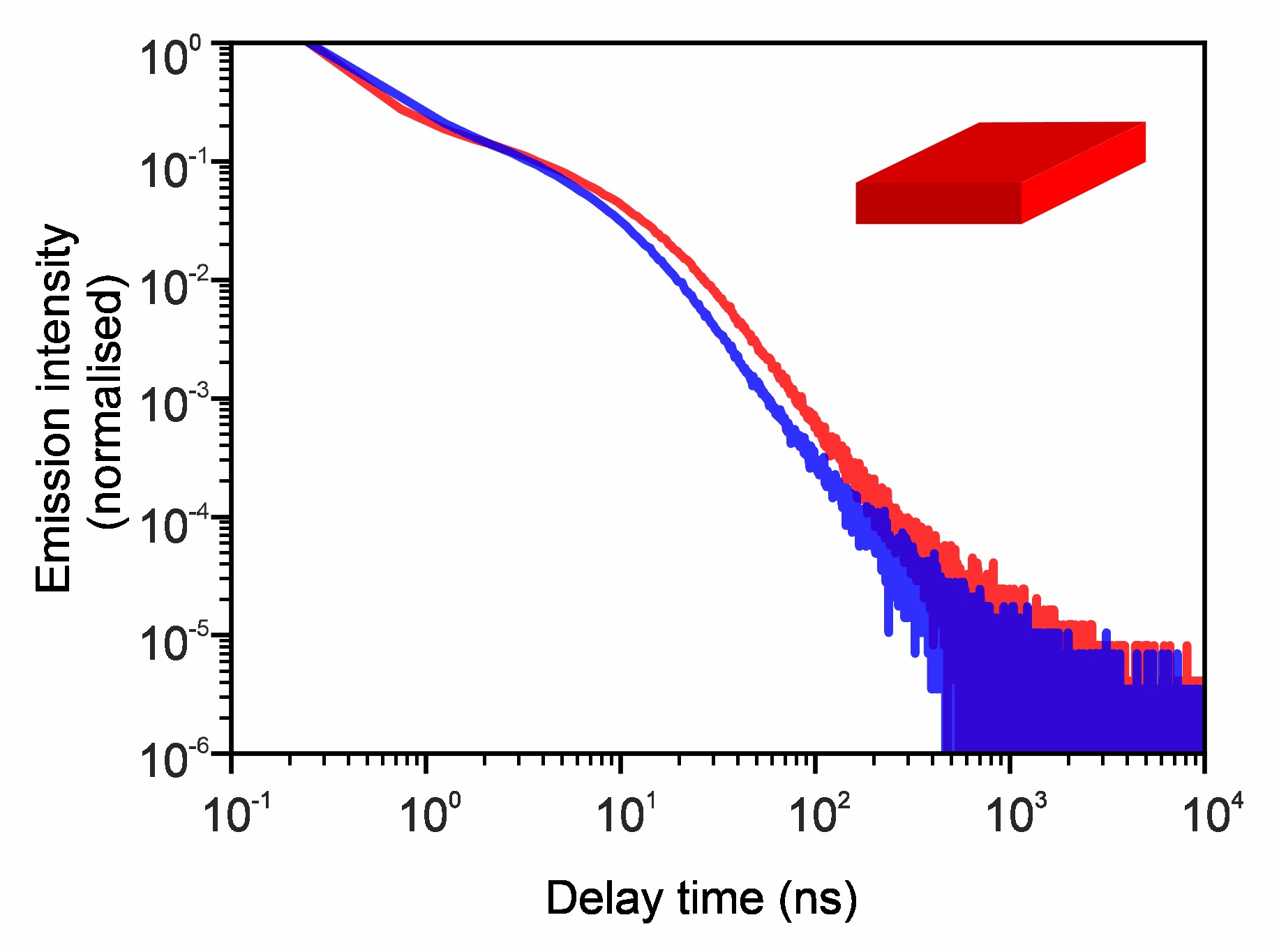}%
\caption{Comparison of curves for photoluminescence intensities for 2D (red) and 3D (blue) models. The core case is shown with a density of defects 1 per 225 surface lattice sites. $\mu=0.8,N_{sim}=10^6, t_{\mathrm{bin}}=500 ps$.}
\label{2D3D_PL2}
\end{figure}

\begin{figure}
\includegraphics[width=8cm]{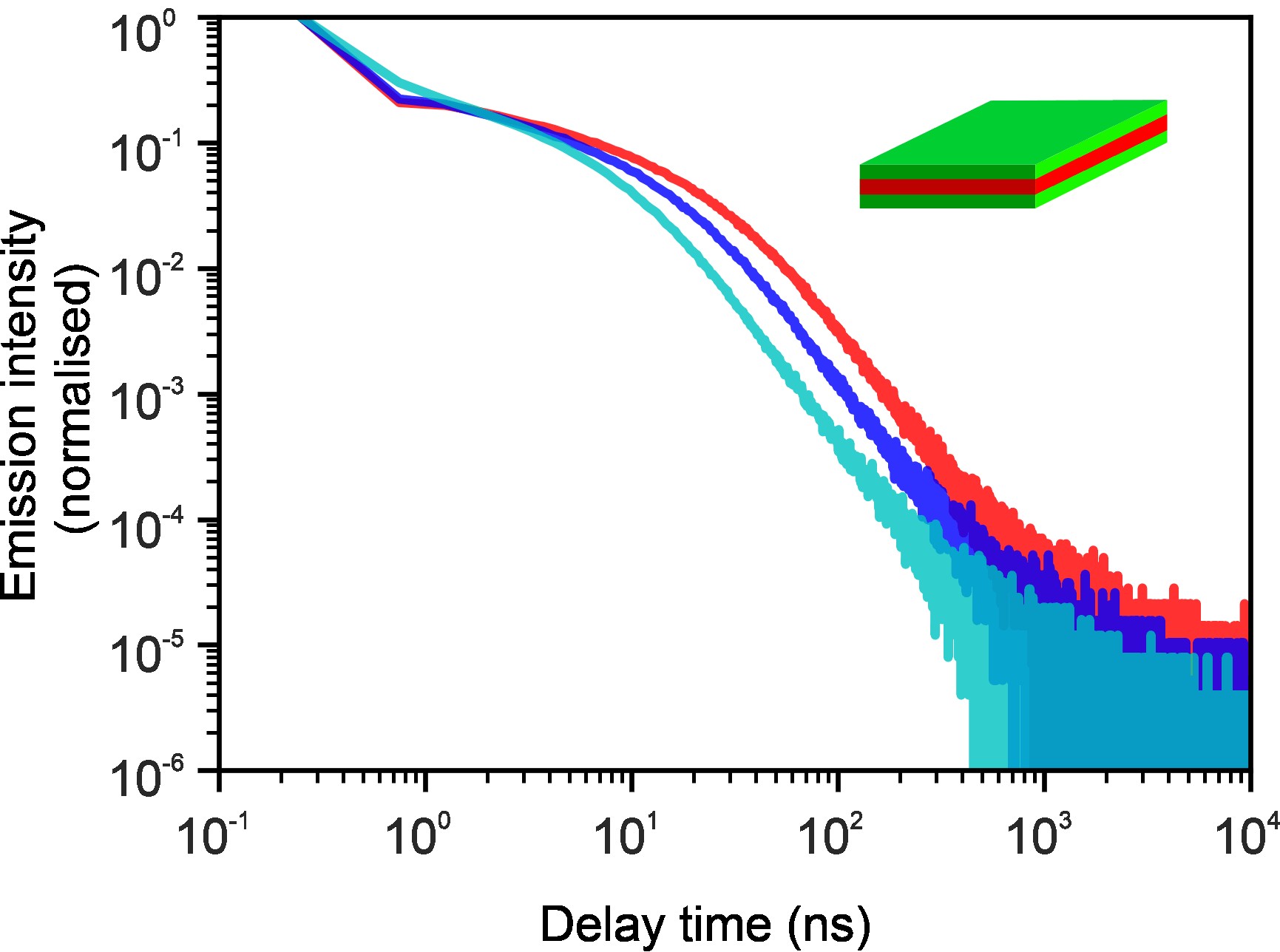}%
\caption{Comparison of curves for photoluminescence intensities for 2D (red), 3D homogeneous (blue) and 3D distinct domains (cyan) models. The core-shell case is shown with a density of defects 1 per 100 surface lattice sites. $\mu=0.8$, $N_{\textrm{sim}}=10^6, t_{\mathrm{bin}}=500 ps$. The kink at short times occurs due to binning. In the first two cases the jumps are considered to be equally likely for all directions. For the last case (cyan data) the layers are split between two regions. The outer region is one material and the inner region is another one. Thus, it corresponds to the core-shell structure more realistically. The jumps from the surface layer into the bulk layers and back are 5 times smaller than the probabilities of jumps within the same region.}
\label{2D3D_PL1}
\end{figure}

For the 3D model the half of NPL vs. the symmetry plane parallel to the surface was simulated with the $N_{l}$ layers ($N \times N \times N_{l}$). The traps are located only in the surface layer. Periodic boundary conditions are applied to in-plane coordinates. The initial exciton positions are generated from discrete uniform distributions along all three coordinates. The exciton cannot escape to the solution so the possible jumps on the surface layer ($N, N, 0$) are $[(+1, 0, 0), (-1, 0, 0), (0, +1, 0), (0, -1, 0), (0, 0, +1)]$ with the equal probabilities of $\frac{1}{5}$. In the internal layer the possible jumps are $[(+1, 0, 0), (-1, 0, 0), (0, +1, 0), (0, -1, 0), (0, 0, +1), (0, 0, -1)]$ with equal probabilities of $\frac{1}{6}$ In order to account for symmetry of NPL, the crossing of the symmetry plane $(0, 0, +1)$ is replaced by $(0, 0, 0)$ for even $N_{l}$ ($2N_{l}$ layers in NPL) or by $(0,0, -1)$ for the odd $N_{l}$ ($2(N_{l}-1) + 1$ layers in NPL).

In Fig. \ref{2D3D_PL2} we see that for core NPLs 2D model and the model of diffusion in a few parallel atomistic layers give the same results. In this case we assume the density of surface defects the same as it was in Fig. \ref{SimulationVsExpCore}, i.e. 1 defect per 225 surface lattice sites. Fig. \ref{2D3D_PL1} shows comparison of three different models. Namely, the 2D model, the 3D model which assumes the same properties and jump probabilities in all 5 layers and the 3D model, where jump probabilities differ when the jumps occur from the surface layers into the bulk. The latter case is the easiest generalisation of a 3D model. It assumes that the jumps into the bulk and back to the surface are 5 times less likely than the jumps within the same layer. The density of defects corresponds to Fig. \ref{SimulationVsExpCoreShell}, i.e. 1 per 100 surface lattice sites. All three curves show the same tail behaviour as one would expect. The transitional region changes its shape and shifts, i.e. more complex models allow for finer description. We see that the available experimental data can be fitted well with a relatively simple 2D approach which tells us that 2D model is sufficient as a minimal model.

\section*{Appendix D: Proof of the non-negativity of Eq. (\ref{Equations}) solutions}

The concentration of excitons must be non-negative. Hence, according to the Bernstein theorem \cite{Feller}, its Laplace transform $\tilde{n}_f(s)$ should be completely monotonic (c.m.). The function $\tilde g(s)$ is called c.m. if $(-1)^n \tilde g^{(n)}(s)\ge1$ for $s>0$ and $n=0,1,2...$. If $m_1(s)$ and $m_2(s)$ are completely monotonic then their product $m_1(s)m_2(s)$ is also c.m. Also if $\tilde f(s)$ is c.m. and $\tilde h(s)$ is Bernstein function (i.e. $\tilde h\ge0$ for positive $s$ and $\tilde h'(s)$ is c.m.) then $\tilde f(\tilde h(s))$ is c.m. \cite{Feller}.

In order to show the complete monotonicity of $\tilde n_f(s)$ we present it as $\tilde n_f(s)=\tilde f(\tilde h(s))$, where $\tilde f=1/s$ and $\tilde h(s)=s+\alpha+\beta(1-\tilde\gamma(s))$. The function $\tilde h(s)$ is positive since $\tilde \gamma(s)$ is the Laplace transform of the probability distribution function and, therefore, $\tilde \gamma(s)\le1$. The function $\tilde\gamma(s)$ is c.m. as the Laplace transform of the PDF, i.e. $(-1)^n\tilde\gamma^{(n)}(s)\ge0$. The derivative $\tilde h'(s)=1-\beta\tilde\gamma'(s)$ is c.m. since $\tilde\gamma(s)$ is c.m. Thus, due to the property mentioned above $\tilde n_f(s)=\tilde f(\tilde h(s))$ is c.m. as well and $n_f(t)$ is non-negative due to the Bernstein theorem \cite{Feller}.

\section*{Appendix E: The derivation of the limit result for the two state Markovian limit}

We show here how one can recover the limit of a two state Markovian system with constant rates of exchange and a decay in one of the states (considered, for instance, in Van Kampen's book \cite{VanKampen}, Chapter VII, section 5, exercise (5.8)). This corresponds to the Markovian limit of Eqs. (\ref{Equations})-(\ref{Equations1}).

The case of two states with exchange can be written as
\begin{eqnarray}\label{VanKampenLimit}
&&\frac{dn_f(t)}{dt}=-\alpha n_f(t)-\beta n_f(t)+n_t(t)/\tau_*,\\
&&\frac{dn_t(t)}{dt}=\beta n_f(t)-n_t(t)/\tau_*,
\label{VanKampenLimit0}
\end{eqnarray}
where $n_f(t)$ and $n_t(t)$ correspond to the  free and trapped states in our model and $\alpha$, $\beta$ and $\tau_*$ are positive constants.
 
Let us treat Eq. (\ref{VanKampenLimit0}) formally as a non-homogeneous ordinary differential equation of the first order. Then its formal solution reads
\begin{equation}
n_t(t)=\int_0^t dt' e^{-(t-t')/\tau_*}\beta n_f(t').    
\end{equation}
By substituting  the last equation into (\ref{VanKampenLimit}) we get
\begin{equation}
\frac{dn_f(t)}{dt}=-\alpha n_f(t)-\beta n_f(t)+\frac{1}{\tau_*}\int_0^t dt' e^{-(t-t')/\tau_*}\beta n_f(t') \label{Markovnf}
\end{equation}
and, respectively,
\begin{equation}
\frac{dn_t(t)}{dt}=\beta n_f(t)-\frac{1}{\tau_*}\int_0^t dt' e^{-(t-t')/\tau_*}\beta n_f(t'). 
\label{Markovnt}
\end{equation}

We see that the system of equations (\ref{Markovnf})-(\ref{Markovnt}) is a particular case of our model Eqs. (\ref{Equations})-(\ref{Equations1}) with exponential trapping PDF $\gamma(t)=\frac{1}{\tau_*}\exp\left(-t/\tau_*\right)$.
Thus, we have established a link of the non-Markovian model (\ref{Equations})-(\ref{Equations1}) and the classical Markovian two-state equations \cite{VanKampen}.

\section*{Appendix F: Solution of Markovian kinetic rate equations}

We consider equations (\ref{VanKampenLimit}), (\ref{VanKampenLimit0}) with initial conditions $n_f(t=0)=N_0$, $n_t(t=0)=0$. In the Laplace space 
\begin{eqnarray}
 &&s\tilde{n}_f-N_0=-\alpha \tilde{n}_f-\beta \tilde{n}_f+\tilde{n}_t/\tau_*,\\
 &&s\tilde{n}_t=\beta \tilde{n}_f- \tilde{n}_t/\tau_*
\end{eqnarray}
or,
\begin{eqnarray}
 &&(s+\alpha+\beta)\tilde{n}_f-\tilde{n}_t/\tau_*=N_0,\\
 &&-\beta \tilde{n}_f +(s+1/\tau_*)\tilde{n}_t=0.
\end{eqnarray}
Determinant
\begin{eqnarray}
Det=s^{2}+(\alpha+\beta+1/\tau_*)s+\alpha/\tau_*.
\end{eqnarray}
Solution reads
\begin{eqnarray}\label{marknf}
 &&\tilde{n}_f(s)=N_0\frac{s+1/\tau_*}{s^{2}+(\alpha+\beta+1/\tau_*)s+\alpha/\tau_*},\\
 &&\tilde{n}_t(s)=N_0\frac{\beta}{s^{2}+(\alpha+\beta+1/\tau_*)s+\alpha/\tau_*}.
 \label{marknt}
\end{eqnarray}
The denominator can be written as
\begin{eqnarray}
s^{2}+(\alpha+\beta+1/\tau_*)s+\alpha/\tau_*=(s+|s_1|)(s+|s_2|),
\label{quad}
\end{eqnarray}
where $s_1$ and $s_2$ are the roots of the quadratic equation 
\begin{eqnarray}
s^{2}+(\alpha+\beta+1/\tau_*)s+\alpha/\tau_*=0,
\end{eqnarray}
that is
\begin{eqnarray}\nonumber
 &&s_1=-\frac{\alpha+\beta+1/\tau_*}{2}+\sqrt{\frac{1}{4}(\alpha+\beta+1/\tau_*)^{2}-\alpha/\tau_*}<0,\\\nonumber
 &&s_2=-\frac{\alpha+\beta+1/\tau_*}{2}-\sqrt{\frac{1}{4}(\alpha+\beta+1/\tau_*)^{2}-\alpha/\tau_*}<0.
\end{eqnarray}

After plugging (\ref{quad}) into (\ref{marknf}),(\ref{marknt}) we get
\begin{eqnarray}\nonumber
 &&\tilde{n}_f(s)=N_0\frac{s+1/\tau_*}{(s+|s_1|)(s+|s_2|)}=
 \\&&\frac{N_0}{|s_2|-|s_1|}\left\{\frac{1/\tau_*-|s_1|}{s+|s_1|}+\frac{|s_2|-1/\tau_*}{s+|s_2|}\right\},\\\nonumber
 &&\tilde{n}_t(s)=N_0\frac{\beta}{(s+|s_1|)(s+|s_2|)}=
 \\&&\frac{N_0\beta}{|s_2|-|s_1|}\left\{\frac{1}{s+|s_1|}-\frac{1}{s+|s_2|}\right\}.
\end{eqnarray}

Taking inverse Laplace transform
\begin{eqnarray}\nonumber
&&n_f(t)=\frac{N_0}{|s_2|-|s_1|}\left\{(1/\tau_*-|s_1|)e^{-|s_1|t}+\right.\\\label{marksolnf}
&&\left.(|s_2|-1/\tau_*)e^{-|s_2|t} \right\},\\
&&n_t(t)=\frac{N_0\beta}{|s_2|-|s_1|}\left\{e^{-|s_1|t}-e^{-|s_2|t} \right\}.\label{marksolnt}
\end{eqnarray}

\textbf{Remark 1.} It is possible to prove that $|s_1|<1/\tau_*<|s_2|$, thus both terms in braces of (\ref{marksolnf}) are positive and, automatically, the term in braces of (\ref{marksolnt}) is also positive. For example,
\begin{eqnarray}
1/\tau_*>|s_1|=\frac{\alpha+\beta+1/\tau_*}{2}-\sqrt{\frac{1}{4}(\alpha+\beta+1/\tau_*)^{2}-\alpha/\tau_*},
\end{eqnarray}
\begin{eqnarray}
\sqrt{\frac{1}{4}(\alpha+\beta+1/\tau_*)^{2}-\alpha/\tau_*}>\frac{\alpha+\beta+1/\tau_*}{2}-1/\tau_*.
\end{eqnarray}

If $\alpha+\beta<1/\tau_*$, then the left side $>0$, right side $<0$ .
If $\alpha+\beta>1/\tau_*$, then taking square of both sides, 
\begin{eqnarray}\nonumber
&&\frac{1}{4}(\alpha+\beta+1/\tau_*)^{2}-\alpha/\tau_*>\\\nonumber
&&\frac{1}{4}(\alpha+\beta+1/\tau_*)^{2}-1/\tau_*(\alpha+\beta+1/\tau_*)+1/\tau_*^{2}\Rightarrow\\\nonumber
&&0>-\beta/\tau_*.
\end{eqnarray}

In the same way the inequality $1/\tau_*<|s_2|$ is proved.

\textbf{Remark 2.} If $\alpha=0$, then $s_1=0$, $s_2=-(\beta+1/\tau_*)$, then $n_{f}(t)$ and $n_t(t)$ reduce to 
\begin{eqnarray}
 &&n_f(t)=\frac{N_0}{\beta+1/\tau_*}\left\{1/\tau_*+\beta e^{-(\beta+1/\tau_*)t} \right\},\\
 &&n_t(t)=\frac{N_0\beta}{\beta+1/\tau_*}\left\{ 1-e^{-(\beta+1/\tau_*)t} \right\}.
\end{eqnarray}

\textbf{Remark 3. Normalisation.} $\tilde{n}_{f}(s=0)=N_0/\alpha=const$. This implies normalisation $\int_{0}^{\infty}n_{f}(t)dt=N_0/\alpha$ what follows directly from Eq. (\ref{exactsolutionfree}).

\section*{Appendix F: Derivation of the asymptotic power-law for $n_f(t)$}\label{asymptotic}

In order to derive Eq. (\ref{intensity}) for the emission intensity in the long-time limit we follow the reasoning from \cite{HavlinWeiss1990} and use an identity
\begin{equation}
\frac{N_0}{\alpha}-\tilde n_f(s)=\int_0^\infty \left[1-e^{-st}\right]n_f(t)dt.
\label{identity}
\end{equation}
From the expansion of $\tilde n_f(s)$ at small $s$, Eq. (\ref{limitnf}), we infer that $n_f(t)$ decays at long times as $n_f(t\gg\tau_*)\simeq\frac{C\tau_*^\mu}{t^{1+\mu}}$. We put the latter expression in the identity and get
\begin{equation}
\frac{N_0}{\alpha}-\tilde n_f(s)\simeq C\int_{T_0}^\infty \left[1-e^{-st}\right]\frac{\tau_*^\mu}{t^{1+\mu}}dt.    
\end{equation}
Now we take derivatives from both sides with respect to $s$ and using Tauberian theorem for Laplace transforms \cite{Feller}, we get
\begin{equation}
-\tilde n_f'(s)\simeq C\Gamma(1-\mu)\tau_*^\mu s^{\mu-1}.   
\end{equation}
After differentiating Eq. (\ref{limitnf}) one obtains the expression for $C$,
\begin{equation}
C=\frac{\mu\beta N_0}{\alpha^2\Gamma(1-\mu)}.    
\end{equation}

\bibliography{literature}

\end{document}